\documentclass[runningheads]{llncs}
\usepackage{graphicx}

\usepackage{tikz}
\usepackage{comment} 
\usepackage{amsmath,amssymb} 
\usepackage{color}

\begin{document}
\pagestyle{headings}
\mainmatter
\def\ECCVSubNumber{6639}  

\title{ByeGlassesGAN: Identity Preserving Eyeglasses Removal for Face Images} 

\titlerunning{ByeGlassesGAN}
%
\author{Yu-Hui Lee\inst{1}\orcidID{0000-0002-4162-1597} \and
Shang-Hong Lai\inst{1,2}}
\authorrunning{Y. Lee and S. Lai}
%
\institute{Department of Computer Science, National Tsing Hua University, Taiwan \and
Microsoft AI R\&D Center, Taiwan
\\
\email{s106062508@m106.nthu.edu.tw},
\email{shlai@microsoft.com}}
\maketitle

\begin{abstract}
In this paper, we propose a novel image-to-image GAN framework for eyeglasses removal, called ByeGlassesGAN, which is used to automatically detect the position of eyeglasses and then remove them from face images. Our ByeGlassesGAN consists of an encoder, a face decoder, and a segmentation decoder. The encoder is responsible for extracting information from the source face image, and the face decoder utilizes this information to generate glasses-removed images. The segmentation decoder is included to predict the segmentation mask of eyeglasses and completed face region. The feature vectors generated by the segmentation decoder are shared with the face decoder, which facilitates better reconstruction results. Our experiments show that ByeGlassesGAN can provide visually appealing results in the eyeglasses-removed face images even for semi-transparent color eyeglasses or glasses with glare. Furthermore, we demonstrate significant improvement in face recognition accuracy for face images with glasses by applying our method as a pre-processing step in our face recognition experiment.
\end{abstract}

\section{Introduction}

Face recognition has been researched extensively and widely used in our daily lives. Although state-of-the-art face recognition systems are capable of recognizing faces for practical applications, their accuracies are degraded when the face images are partially occluded, such as wearing eyeglasses. An obvious reason causes this problem is that the eyeglasses may occlude some important information on faces, leading to discrepancies in facial feature values. For example, the thick frame of glasses may block the eyes. Hence, in the past, researchers proposed to apply the PCA-based methods~\cite{wu2004automatic,park2005glasses} to remove eyeglasses from face images. However, the PCA-based method can only provide approximate glasses removal image via face subspace projection. In addition, they did not really evaluate their methods on diverse face recognition tasks.

\begin{figure}[htb]
\setlength{\abovecaptionskip}{0.cm}
\begin{center}
   \includegraphics[width=0.65\linewidth]{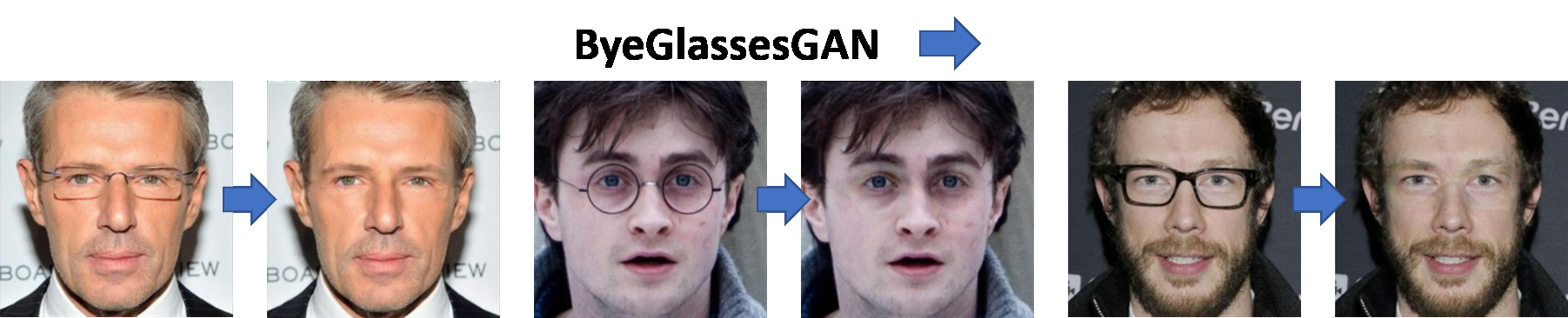}
\end{center}
   \caption{Examples of glasses removal by ByeGlassesGAN.}
\label{fig:first}
\end{figure}

Another reason for the degradation of face recognition accuracy with eyeglasses is that face images with eyeglasses are considerably fewer than glasses-free images. It is hard to make the recognition model learn the feature of various kinds of eyeglasses. Recently, alongside with the popularity of face attributes manipulation, some GAN based methods, such as \cite{guo2018face} and \cite{zhang2018generative}, improved the capability of recognizing faces with eyeglasses by synthesizing a large amount of images of faces with eyeglasses for training a face recognition model.

Different from the previous works, we aim at improving face recognition accuracy by removing eyeglasses with the proposed GAN model before face recognition. With the proposed GAN-based method, we can not only improve face recognition accuracy, the visually appealing glasses-removed images can also be used for some interesting applications, like applying virtual makeup. 

The main contributions of this work are listed as follows:
\begin{enumerate}
\item We propose a novel glasses removal framework, which can automatically detect and remove eyeglasses from a face image.
\item Our proposed framework combines the mechanisms of the feature sharing between 2 decoders to acquire better visual results, and an identity classifier to make sure the identity in the glasses-removed face image is well preserved. 
\item We come up with a new data synthesis method to train a glasses removal network, which effectively simulates color lens, glare of reflection as well as the refraction on eyeglasses. 
\item In the experiment, we demonstrate that the face recognition accuracy is significantly improved for faces with eyeglasses after applying the proposed eyeglasses removal method as a pre-processing step.   
\end{enumerate}

\section{Related Works}

\subsection{Face Attributes Manipulation}
Face attributes manipulation is a research topic that attracts a lot of attention. Along with the popularity of GAN, there are many impressive GAN-based methods proposed for editing face attributes. \cite{li2016deep} and \cite{zhang2018generative} edit face attributes through an attribute transformation network and a mask network. Both of them preserve the identity of the source images by using the predicted mask to constrain the editing area. AttGAN~\cite{he2017arbitrary} edits face images through the attribute classification constraint and reconstruction learning. ELEGANT~\cite{xiao2018elegant} can not only manipulate face images but also manipulate images according to the attributes of reference images. ERGAN~\cite{hu2019unsupervised} removes eyeglasses by switching features extracted from a face appearance encoder and an eye region encoder. Besides, there are several face attributes editing methods which are not GAN-based. For example, DFI~\cite{upchurch2017deep} manipulated face images through linear interpolation of the feature vectors of different attributes. \cite{bao2018towards} achieved identity preserving face attributes editing by disentangling the identity and attributes vectors of face images with the mechanisms of Variational Autoencoder and GAN.
However, these face attributes manipulation methods suffer from the instability problem. For example, the identity may not be preserved or some artifacts may be generated.

\subsection{Image Completion}

Eyeglasses removal can also be seen as a face image completion problem. 
Recently there are many deep learning works~\cite{pathak2016context,yu2018generative,iizuka2017globally,yang2017high,liu2018image,jo2019sc} focusing on image completion. Context Encoder~\cite{pathak2016context} is the first deep learning and GAN based inpainting method. After that, \cite{iizuka2017globally} significantly improved the quality of inpainting results by using both a global and a local discriminators, with one of them focusing on the whole image and the other focusing on the edited region. PartialConv~\cite{liu2018image} masks the convolution to reduce the discrepancy (e.g. color) between the inpainted part and the non-corrupted part. Recently, there are also some interesting and interactive completion methods~\cite{yu2018free,jo2019sc} which support free-form input. Users can easily control the inpainted result by adding desired sketches on the corrupted regions.

The main difference between the proposed ByeGlassesGAN and existing image completion methods is that our method does not require a predefined mask for the completion. In fact, the eyeglasses removal problem is not the same as image completion because the glasses region could be either transparent or semi-transparent. Our method can exploit the original image in the glasses region to provide better glasses-removed result. Besides, compared with the face attributes manipulation methods described above, our method can automatically remove the glasses and better preserve the face identity in the glasses-removed images.

\section{ByeGlassesGAN}\label{c:methods}

In this paper, we propose a multi-task learning method which aims at predicting the position of eyeglasses and removing them from the source image. Since we expect the eyeglasses-removed images can improve the performance of face recognition, the generated results of ByeGlassesGAN must look realistic and well preserve the identities of the source images.

\begin{figure}
\setlength{\abovecaptionskip}{0.cm}
\begin{center}
\includegraphics[width=0.85\linewidth]{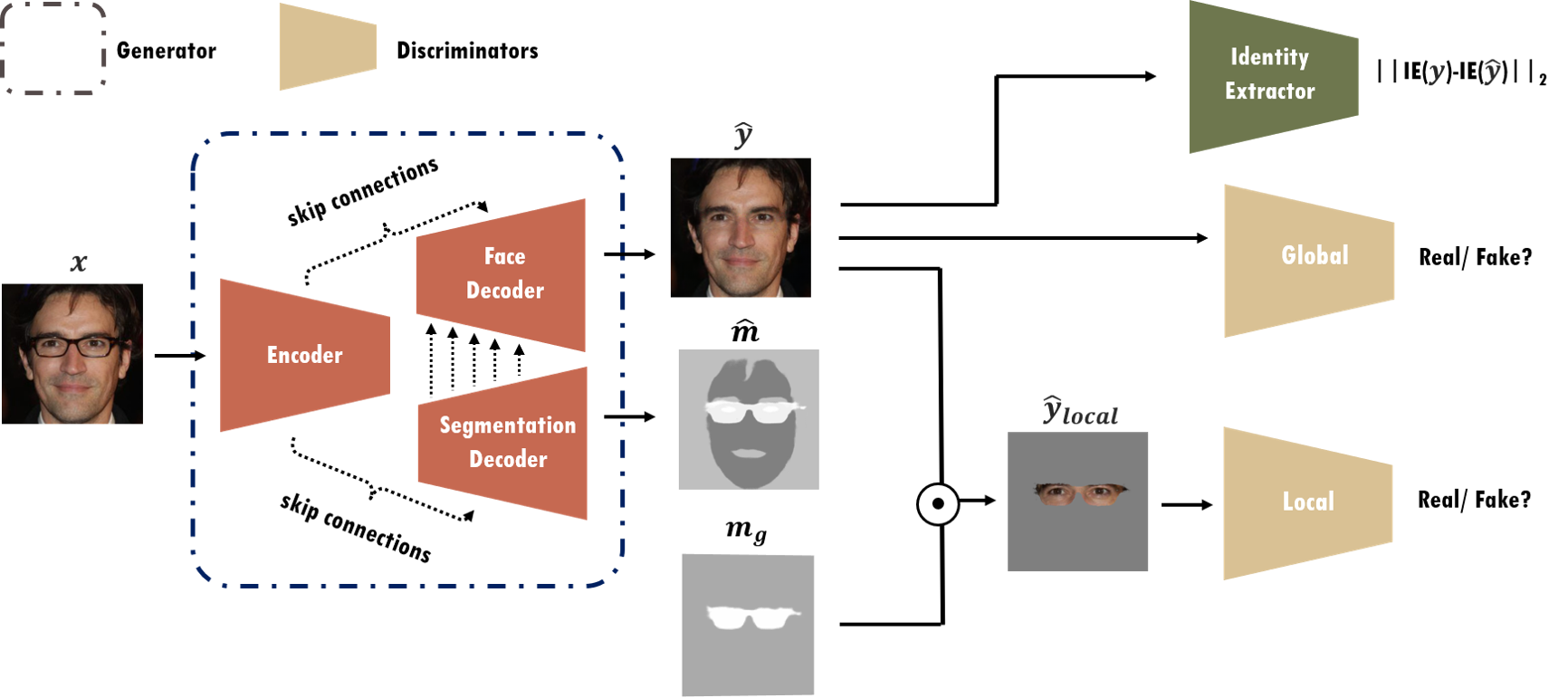}
\end{center}
   \caption{\textbf{Framework of ByeGlassesGAN.} Each input image ($x$) is first fed into the Encoder to encode the feature vector. The Face Decoder and the Segmentation Decoder then manipulate the glasses-removed image ($\hat{y}$) and the segmentation mask ($\hat{m}$) of eyeglasses and face shape with the extracted vectors. Two discriminators are included to make sure both the whole generated image ($\hat{y}$) and the edited part ($\hat{y}_{Local}$) look realistic. An Identity Extractor is also applied to minimize the distance between the identity feature vectors computed from the output image ($\hat{y}$) and the ground truth image.}
\label{fig:struct}
\end{figure}

\subsection{Proposed Framework} 

Figure~\ref{fig:struct} illustrates the overall framework of our ByeGlassesGAN, which contains a generator, an identity extractor, and two discriminators. The generator (G) can be separated into 3 deep neural networks, encoder (E), face decoder (FD), and segmentation decoder (SD). Here we assume the training data contains a set of face images ($x$) associated with the corresponding glasses-removed images ($y$) and the corresponding masks ($m$) of eyeglasses region and the completed face shape. Given a source face image $x$, which first goes through the encoder to encode the feature vector of image $x$. After that, we synthesize the glasses-removed image $\hat{y}$ with face decoder using the feature vector mentioned above. Meanwhile, a segmentation decoder is there for generating the binary mask $\hat{m}$ of the glasses region. However, after testing with this baseline model, we found that although there are many good removal results, when the eyeglasses are special or the face is not frontal, the removal effect may degrade. Hence, we were wondering whether there exists a good representation for face that can help remove eyeglasses. Since eyeglasses removal can be regarded as a kind of inpainting task on the face region,  we can include semantic segmentation mask of face and eyeglasses regions into the framework. The segmentation of face shape is an excellent hint to guide FD to know the characteristics of each pixel in the face region and should maintain consistency with the neighboring pixels. After the experiment, we found that making SD predict the binary mask of face shape as well greatly improves the glasses-removed results. Hence, we let SD not only predict the binary mask of eyeglasses, but also the mask of face shape. Besides, the information obtained from SD is shared with FD with the skip connections to guide FD synthesizing images. Thus, we have

\begin{equation}
\hat{y}=FD(E(x))
\end{equation}
and
\begin{equation}
\hat{m}=SD(E(x))
\end{equation}
where $\hat{m}$ is a 2-channel mask, one of the channels indicates the position of the glasses region, and the other is for the face shape.
Furthermore, in order to ensure the quality of synthetic output $\hat{y}$, we adopt a global and a local discriminator~\cite{iizuka2017globally} to make sure both the synthetic image $\hat{y}$ and the inpainted frame area $\hat{y}_{Local}$ look realistic. Besides, we also include an Identity Extractor to minimize the distance between the identity feature vectors computed from the output image ($\hat{y}$) and the ground truth image ($y$).

\subsection{Objective Function}

The proposed ByeGlassesGAN is trained with the objective function consisting of four different types of loss functions, i.e. the adversarial loss, per-pixel loss, segmentation loss, and identity preserving loss. They are described in details subsequently.

\subsubsection{Adversarial Loss}

In order to make the generated images as realistic as possible, we adopt the strategy of adversarial learning. Here we apply the objective function of LSGAN~\cite{mao2017least,zhu2017unpaired} since it can make the training process of GAN more stable than the standard adversarial loss. 
Here we adopt 2 kinds of GAN loss, $L_{GAN}^{Global}$ and $L_{GAN}^{Local}$ for training the discriminators. Equation~\ref{dgan} shows the global adversarial loss $L_{D}^{Global}$.

\begin{equation}
L_{D}^{Global}= \mathbb{E}_{y\sim P_{y}}[(D_{Global}(y)-1)^2] +\mathbb{E}_{x\sim P_{x}}[(D_{Global}(\hat{y}))^2]
\label{dgan}
\end{equation}

When computing $L_{D}^{Local}$, we replace $y$, $\hat{y}$, $D_{Global}$ by $y_{Local}$, $\hat{y}_{Local}$, and $D_{Local}$ in Equation~\ref{dgan}. $y_{Local} = y\odot m_{g}$, and $\hat{y}_{Local} = \hat{y}\odot m_{g}$. $\odot$ denotes the element-wise product operator, and $m_{g}$ is the ground truth binary mask of eyeglasses region.\\
For training the generator, the GAN loss is shown below(Equation~\ref{ggan}). When computing $L_{G}^{Local}$, we also replace $\hat{y}$ and $D_{Global}$ by $\hat{y}_{Local}$ and $D_{Local}$ in Equation~\ref{ggan}.

\begin{equation}
\begin{aligned}
L_{G}^{Global}= \mathbb{E}_{x\sim P_{x}}[(D_{Global}(\hat{y})-1)^2]
\end{aligned}
\label{ggan}
\end{equation}

\subsubsection{Per-pixel Loss}
We compute the $L_1$ distance between the generated image $\hat{y}$ and the ground truth image $y$. Per-pixel loss enforces the output of generator to be similar to the ground truth. We adopt two kinds of $L_1$ loss, $L_{L_1}^{Global}$ and $L_{L_1}^{Local}$. $L_{L_1}^{Local}$ is used for enhancing the removal ability of the generator in the edited region. The global L1 loss is given by 
\begin{equation}
\begin{aligned}
L_{L_1}^{Global}= L1(\hat{y}, y) =  \mathbb{E}_{x\sim P_{x}}[\left \| y-\hat{y} \right \|_{1}]
\end{aligned}
\label{l1}
\end{equation}
When computing $L_{L_1}^{Local}$, we replace $\hat{y}$ and $y$ by $\hat{y}_{Local}$ and $y_{Local}$ in Equation~\ref{l1}.

\subsubsection{Segmentation Loss}
Since we expect ByeGlassesGAN to predict the segmentation mask which facilitates eyeglasses removal, here we adopt binary cross entropy loss for generating the segmentation mask of the eyeglasses region and the face shape. It is given by
\begin{equation}
\begin{aligned}
L_{Seg} = \mathbb{E}_{x\sim P_{x}} -(m\cdot log(\hat{m})+(1-m)\cdot log(1-\hat{m}))
\end{aligned}
\end{equation}
where $\hat{m}$ is the generated mask, and $m$ denotes the ground truth segmentation mask.

\subsubsection{Identity Preserving}\label{ss:IE}
In order to preserve the identity information in the glasses-removed images, we employ an Identity Extractor (IE), which is in fact a face classifier. 

The identity distance loss is introduced to our generator, which is used to minimize the distance between $IE(y)$ and $IE(\hat{y})$. Similar to the concept of perceptual loss, after extracting the feature of $y$ and $\hat{y}$ through the identity extractor, we compute the mean square error between these two feature vectors, given by 
\begin{equation}
\begin{aligned}
L_{ID} = \mathbb{E}_{x\sim P_{x},y\sim P_{y}}[\left \| IE(\hat{y})-IE(y) \right \|_{2}]
\end{aligned}
\end{equation}

This loss encourages the eyeglasses-removed image $\hat{y}$ shares the same identity information with ground truth image $y$ in the feature space of the identity extractor model. 

Note that IE is a ResNet34 classifier pretrained on UMDFaces dataset. When training the Identity Classifier, we transpose the output feature vector of layer4 in ResNet into a 512-dimensional vector, and adopt the ArcFace~\cite{deng2018arcface} loss. 

Finally, the overall loss function of the generator is given as follows:
\begin{equation}
L_{G} =  \lambda_{1}L_{G}^{Global} + \lambda_{2}L_{G}^{Local}
+\lambda_{3}L_{L_{1}}^{Global}+\lambda_{4}L_{L_{1}}^{Local}\\
+\lambda_{5}L_{Seg}+\lambda_{6}L_{ID}
\end{equation}

\subsection{Network Architecture}
Our GAN-based eyeglasses removal framework contains a generator, two discriminators, and an identity extractor. There are one encoder (E) and two decoders (face decoder, FD, and segmentation decoder, SD) in our generator. Following ELEGANT~\cite{xiao2018elegant}, the encoder (E) consists of 5 convolutional blocks, and each block contains a convolutional layer followed by an instance normalization layers and LeakyReLU activation. Both of the face decoder (FD) and the segmentation decoder (SD) consist of 5 deconvolutional blocks and an output block. Each deconvolutional block contains a deconvolutional layer followed by an instance normalization layers and ReLU activation. The output block of the FD is a deconvolutional layer followed by Tanh activation, while the output block of the SD is a deconvolutional layer followed by Sigmoid activation. Since the only area expected to be modified in the source image is the region of eyeglasses, other parts of the image should be kept unchanged. Here we adopt U-NET~\cite{ronneberger2015u} architecture to be the generator of our ByeGlasess-GAN. Skip connections are added to the corresponding layers between E-FD and E-SD. U-Net can considerably reduce the information loss compared with the common encoder-decoder. Besides, skip connections are also added between the corresponding layers of SD and FD, which are used for making the information acquired from SD guide the FD to reconstruct images. The network architecture used for the two discriminators are PatchGAN proposed in pix2pix~\cite{isola2017image}.

\section{Synthesis of Face Images with Eyeglasses}\label{s:expdata}

Since we expect the proposed method can not only remove glasses but also help improve face recognition performance for images with eyeglasses, we need to make sure the detailed attributes (eye shape, eye color, skin color etc.) of the glasses-removed face remain the same as those of the source image. To deal with this problem, the best solution is to collect a large scale of well-aligned images of subjects wearing and without wearing glasses. This kind of paired data is difficult to collect, hence, here we generate the well-aligned paired image data via synthesizing adding eyeglasses onto real face images.

\begin{figure}[tbh]
\setlength{\abovecaptionskip}{0.cm}
\begin{center}
    \includegraphics[width=1\linewidth]{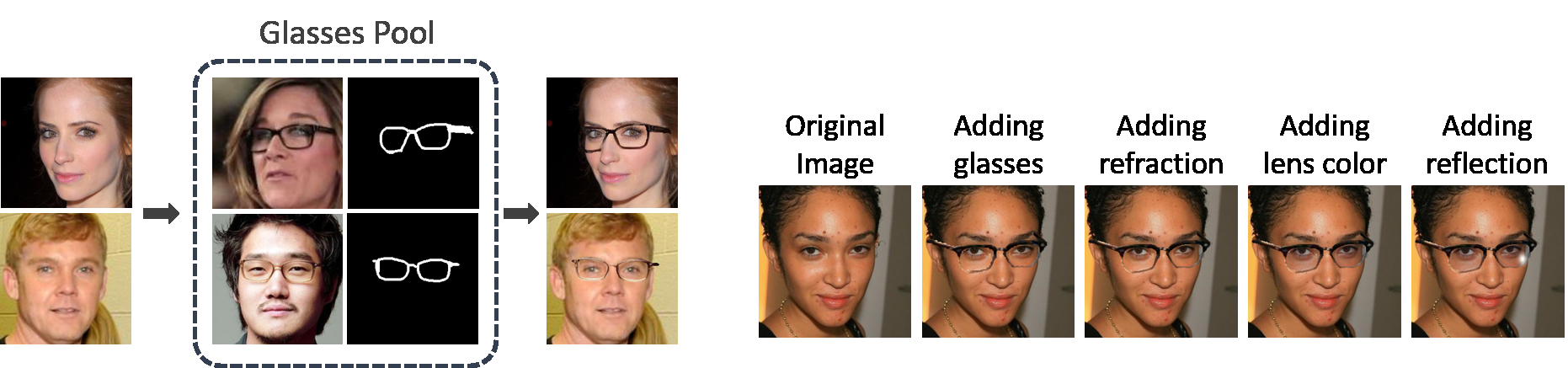}
\end{center}
   \caption{\textbf{How to put on glasses?} We first label the head pose of all the images in our training set, and label 1,000 eyeglasses segmentation mask of the glasses-images to form glasses pool($S_{G}$). Each glasses-free image in the training set can be randomly put on glasses from $S_{G}$ according to the head pose label and the binary mask. After that, we've also applied several photorealism steps to our synthetic images with eyeglasses.}
\label{fig:pool}
\end{figure}

We use CelebA~\cite{liu2015deep} dataset to train the proposed ByeGlassesGAN in our experiments. CelebA is a dataset containing 202,599 images of 10,177 celebrities and annotated with 40 face attribute labels for each image. First, we align all images according to 5 facial landmarks (left/right eye, nose, left/right mouth) into the size of 256x256, and then roughly classify all images into 3 kinds of head pose, frontal, left-front, and right-front using dlib and OpenCV. We manually label the binary masks of 1,000 images with eyeglasses in CelebA as our glasses pool ($S_{G}$), and use the rest of the images with glasses as our testing set. These binary masks precisely locate the position of eyeglasses on each image, so we can make use of them to easily extract 1,000 different styles of glasses. After that, we randomly put glasses from glasses pool onto each glasses-free images according to the head pose (Figure~\ref{fig:pool}, top). In order to make the synthetic glasses images look more realistic to the real one, we randomly apply different levels of deformation around the outer side lens to simulate the refraction of eyeglasses. After that, we dye various colors on the glasses lenses. In addition, we generate many semi-transparent light spot images in different shape. These spots are used to apply on the glasses lenses to simulate the reflected glare on real eyeglasses. This step much improves the ability of our ByeGlassesGAN to manipulate realistic glasses images (Figure~\ref{fig:pool}, bottom). Besides, to generate segmentation mask for the face shape, we used the pre-trained BiSeNet~\cite{yu2018bisenet} which is trained on CelebAMask-HQ~\cite{CelebAMask-HQ} dataset to obtain the face shape mask of the glasses-free images. Finally, we obtain 184,862 pairs of data in total as the training dataset. We will release the glasses pool described above as a new dataset for future research related eyeglasses detection, synthesis, or removal tasks.   

\section{Experimental Results}\label{c:exp_results}

\subsection{Implementation details}
We implement our ByeGlasses GAN with PyTorch. We use Adam for the optimizer, setting $\beta_{1}=0.5$, $\beta_{2}=0.999$, and the learning rate is 0.0002. For the hyperparameters in loss function, we set $\lambda_{1}=1$, $\lambda_{2}=1$, $\lambda_{3}=100$, $\lambda_{4}=200$, $\lambda_{5}=3$, and $\lambda_{6}=5$. Here we train ByeGlassesGAN on a GTX1080 with the batch size set to 16.

\subsection{Qualitative Results}
\begin{figure}[htb]
\setlength{\abovecaptionskip}{0.cm}
\begin{center}
\includegraphics[width=0.75\linewidth]{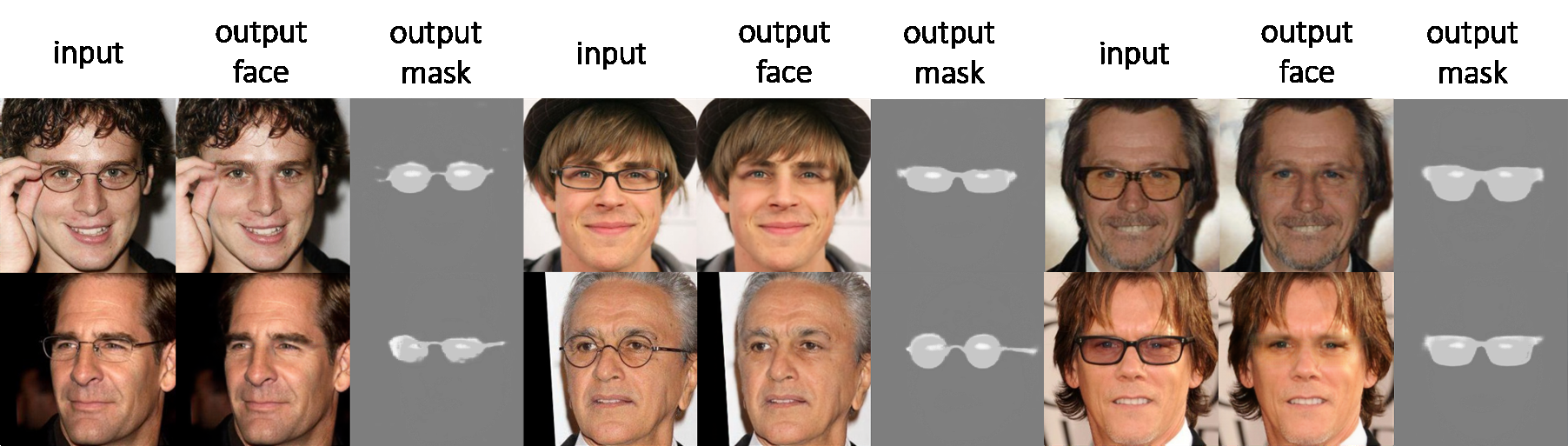}
\includegraphics[width=0.75\linewidth]{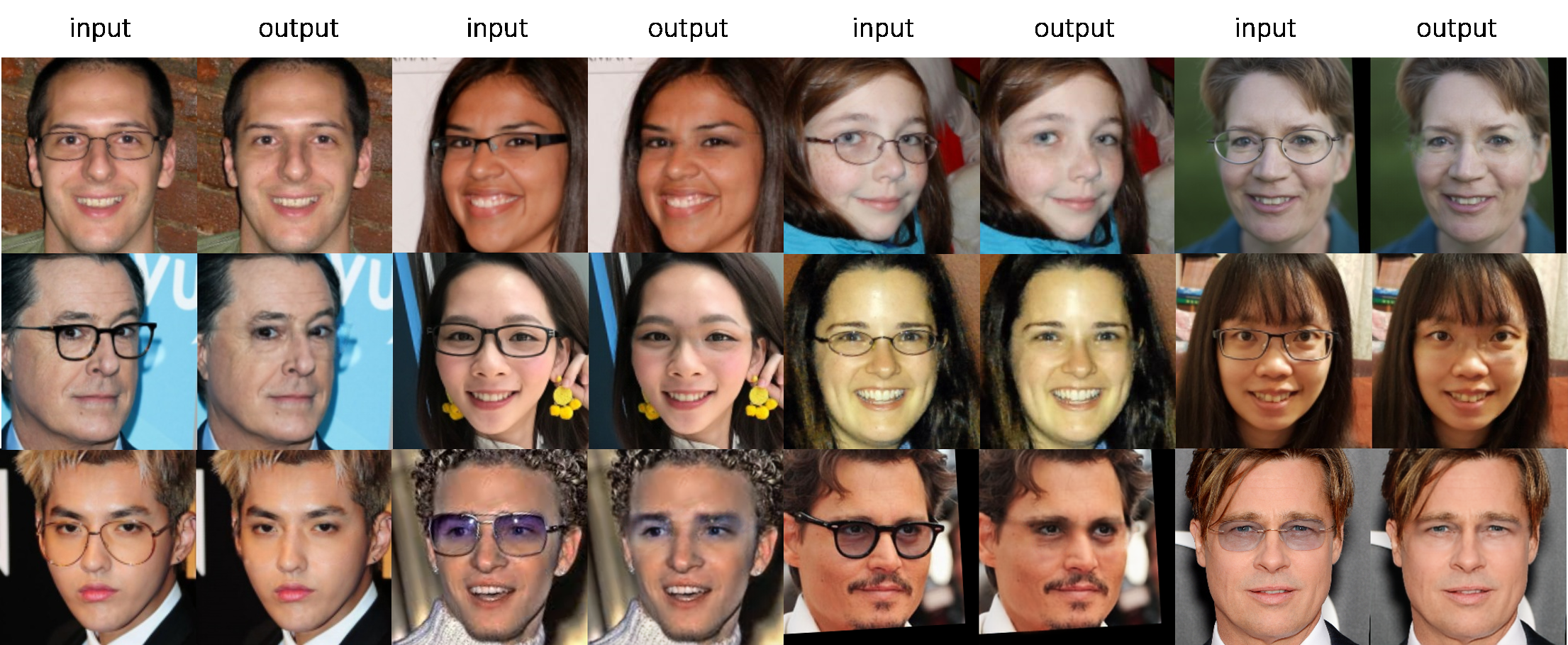}
\end{center}
   \caption{Eyeglasses-removed results. Upper: CelebA results. Bottom: Wild data results. Images credit: MeGlass~\cite{guo2018face} dataset, and photos taken by ourselves.}
\label{fig:result}
\end{figure}

Figure~\ref{fig:result}, Upper shows the visual results of our method on CelebA~\cite{liu2015deep} dataset. All samples are real glasses images in the testing set. The identity of each sample visually remains the same. The generated segmentation masks are also able to point out the accurate region of the face shape. Here we also show some visual results of the face images not in CelebA dataset. Figure~\ref{fig:result}, Bottom shows the visual results of our method testing on the wild data. Our method can not only remove eyeglasses from delicate portraits of celebrities, but also images taken from ordinary camera on mobile phones or laptops. Besides, when we synthesize the training data, we take the head pose of each face image into consideration and generate training image pairs for faces of different poses. Hence, our method is able to deal with non-frontal face images as well. As shown in the bottom row, our method can remove not only the glasses frame but also the tinted lenses.

\begin{figure}[htb]
\setlength{\abovecaptionskip}{0.cm}
\begin{center}
  \includegraphics[width=0.4\linewidth]{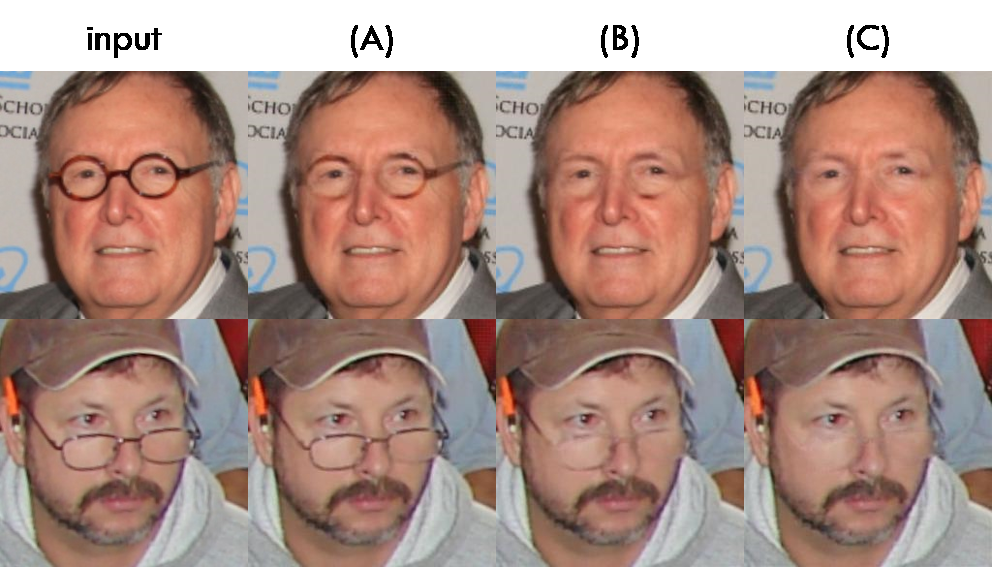}
\end{center}
  \caption{Some glasses removal results under different combinations: (A) baseline model, (B) baseline model with predicting face shape, and (C) complete model in the proposed network.}
\label{fig:ablation}
\end{figure}

Besides, here we also perform experiments on the other two kinds of models which are (A) Baseline model: Segmentation decoder only predicts the binary mask of eyeglasses, and there is no skip connection between FD and SD. (B) Segmentation decoder predicts the binary mask of both eyeglasses and face shape, but there is still no skip connection between FD and SD. As shown in Figure~\ref{fig:ablation}, when we predict the face shape mask in Experiment B, the removal results improve a lot since the face shape mask shares some similar features with the removed result comparing to only predicting glasses mask in Experiment A. Besides, after we add skip connections between the 2 decoders, the segmentation decoder can better guide the face decoder. Sharing the features of the segmentation mask with the face decoder helps the edited region keep consistency with the neighboring skin region, especially when the face is not-frontal or the glasses may not locate in a normal way.

\begin{figure}[htb]
\setlength{\abovecaptionskip}{0.cm}
\begin{center}
   \includegraphics[width=0.8\linewidth]{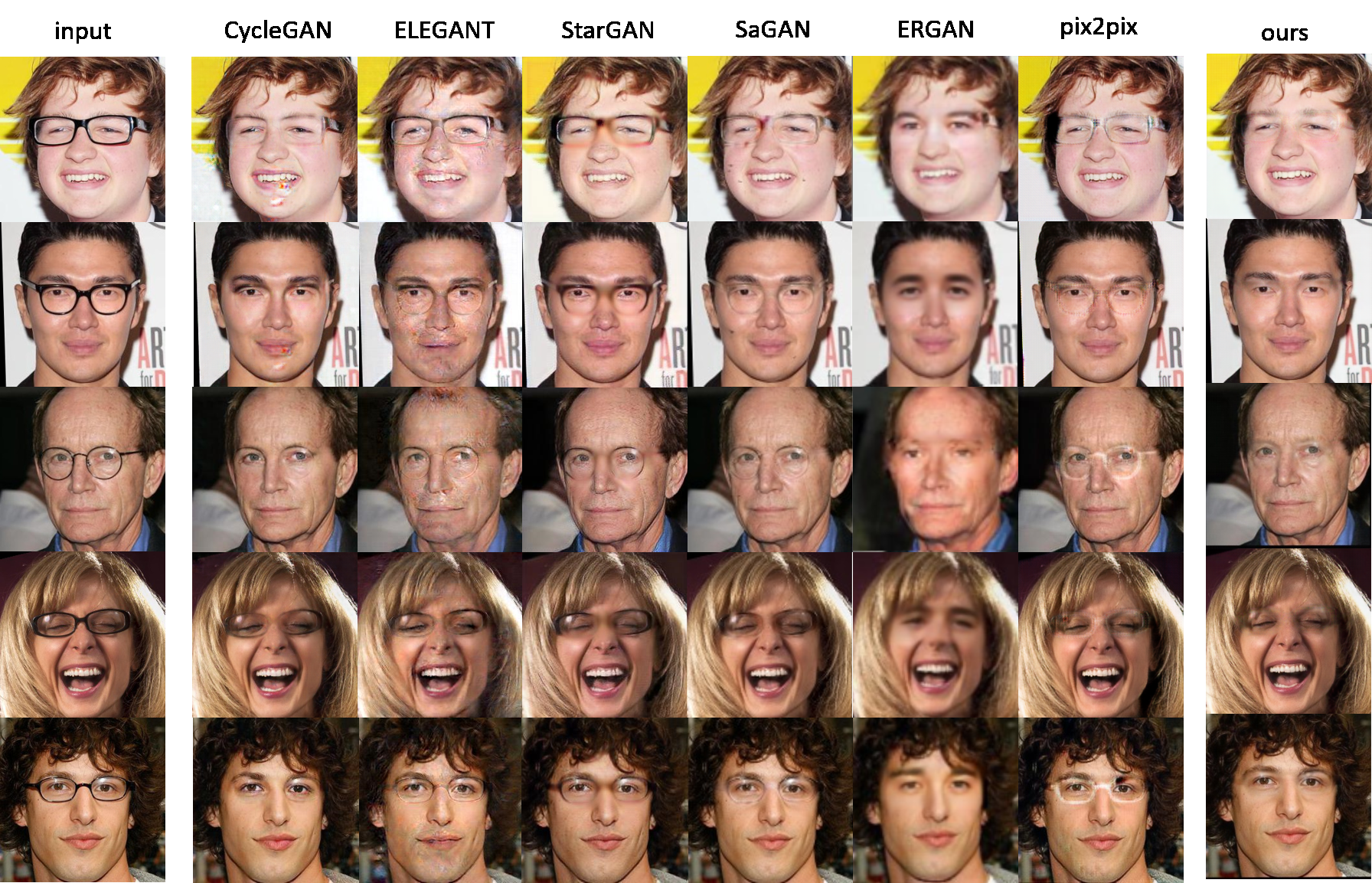}
\end{center}
   \caption{Eyeglasses removal results compared with the other six methods.}
\label{fig:compare}
\end{figure}

We compare the visual results of the proposed method to the other state-of-the-art methods as well, including Pix2pix~\cite{isola2017image}, CycleGan~\cite{zhu2017unpaired}, StarGAN~\cite{choi2018stargan}, ELEGANT~\cite{xiao2018elegant}, ERGAN~\cite{hu2019unsupervised}, and SaGAN~\cite{zhang2018generative}. For comparison, we simply utilize the source code released by the first 5 previous works without any change. For SaGAN, since there is no source code, we carefully implement it ourselves. Pix2pix is a method that needs paired training data, so here we train the pix2pix model using the same data as we mentioned in Section~\ref{s:expdata}. CycleGAN, StarGAN, ELEGANT, and SaGAN are methods adopt unsupervised learning, so we directly use the original CelebA dataset for training the eyeglasses removal networks. ERGAN is an unsupervised method developed for removing eyeglasses. Here we directly apply the model released by the authors to obtain the results. Figure~\ref{fig:compare} shows the removal results of different methods. As shown in the figure, there are many artifacts pop out in pix2pix. For the other 5 unsupervised methods, even there exist visually appealing results, it is still difficult for them to directly remove the glasses stably without generating any artifacts. For ELEGANT and ERGAN, both methods need an additional glasses-free image to guide the removal, so the removal results depend on similarity between the input face image with glasses and the reference image without glasses. Besides, it is worth mentioning that in the last 3 rows in Figure~\ref{fig:compare}, our data synthesis method can effectively strengthen the ability of ByeGlassesGAN to remove reflected glare on the lenses.

\begin{figure}[tb]
\setlength{\abovecaptionskip}{0.cm}
\begin{center}
  \includegraphics[width=0.65\linewidth]{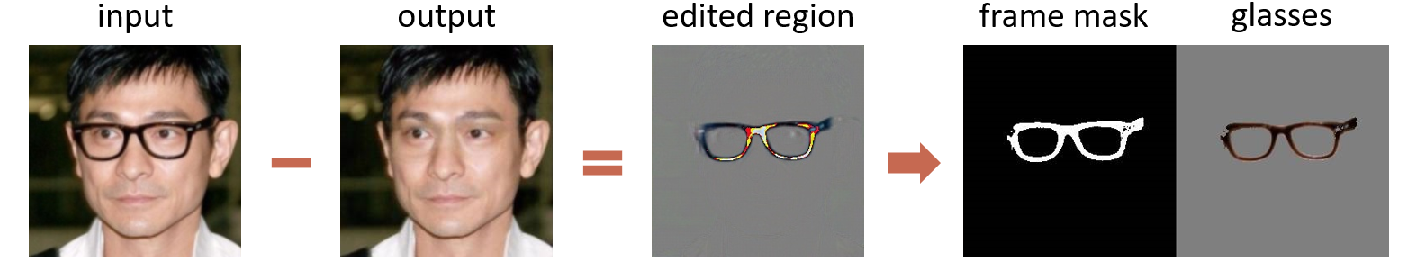}
\end{center}
  \caption{Extracting eyeglasses from the glasses-removed images. These extracted eyeglasses can be used for synthesizing training pairs.}
\label{fig:mkdata}
\end{figure}

Since our method is able to produce high quality glasses-removed result in which the only edited part is the glasses area, we can easily extract the eyeglasses on the input image by applying thresholding operation to the edited region. The edited region is the difference between the input and the output images. Thus, these extracted eyeglasses can also be used for synthesizing training pairs for glasses removal, synthesis, or detection tasks in the future.(Figure~\ref{fig:mkdata})

\subsection{Quantitative Results}

Following \cite{xiao2018elegant}, here we utilize Fr$\Acute{e}$chet Inception Distance(FID)~\cite{heusel2017gans} to measure how well our glasses removal method performs. FID represents the distance between the Inception embedding of the real and the generated images, which reveals how close the embeddings of images in two domains are distributed in feature space. Table~\ref{cfid} shows the FID distances of different methods. Here we perform experiment on 2 different datasets, CelebA~\cite{liu2015deep} and MeGlass~\cite{guo2018face}. The real images set contains real glasses-free images. For the generated images set, it consists of the glasses removed images after applying each of the 6 different models to some real face images with glasses. As shown in Table~\ref{cfid}, our ByeGlassesGAN outperforms the others in both datasets. The FIDs of celebA with CycleGan, StarGAN and ELEGANT methods were reported in \cite{xiao2018elegant}. Besides, we also do the ablation study of removing the segmentation decoder branch, and the perceptual quality of the glasses removal results is not as good as those with the segmentation decoder as shown in Table~\ref{cfid}.

\begin{table}[ht]
\begin{center}
\caption{FID distances of different methods applied on MeGlass and CelebA datasets.}
\resizebox{0.8\linewidth}{!}{
\begin{tabular}{|c|c|c|c|c|c|c|c|c|}
\hline
 & Pix2pix & CycleGAN & StarGAN &  ELEGANT &  SaGAN & ERGAN & ours w/o SD & ours\\
\hline
MeGlass & 39.93 & 29.40 & NULL & 41.09 & 44.94 & 38.25 & 28.26 & \textbf{27.14}\\
CelebA & 50.38 &48.82 & 142.35 & 60.71 & 50.06 & NULL & 44.76 & \textbf{42.97}\\
\hline
\end{tabular}}
\label{cfid}
\end{center}
\end{table}

To further compare our image synthesis method with the others, we conduct a user study. We randomly select 7 portraits with glasses from the testing set and apply different glasses removal methods on all of them, and there are 42 glasses-removed results in total. We invite 49 subjects to evaluate these images and compute the mean opinion score (MOS). As shown in Table~\ref{mos}, apparently, the glasses-removal results by our method are the most preferred since it receives the highest MOS score.

\begin{table}[h]
\begin{center}
\caption{Mean opinion scores of the glasses-removed results of different methods. It is obvious that our ByeGlassesGAN has the highest score.}
\resizebox{0.65\linewidth}{!}{
\begin{tabular}{|c|c|c|c|c|c|c|}
\hline
Methods & Pix2pix & CycleGAN & StarGAN & ELEGANT & SaGAN & Ours\\
\hline
MOS & 2.43 & 3.23 & 2.06 & 1.82 & 2.65 & \textbf{4.31}\\
\hline
\end{tabular}}
\label{mos}
\end{center}
\end{table}

\section{Face Recognition Evaluation}\label{s:face_recog}

In this section, we demonstrate the effect of using our glasses removal GAN as a pre-processing step for the face recognition task.

First, we train a face recognition model on the whole UMDFaces dataset. UMDFaces~\cite{bansal2016umdfaces} is a well-annotated dataset containing 367,888 images of 8,277 subjects, there are both faces with and without eyeglasses. The face recognition model we use here is the PyTorch implementation of MobileFaceNets~\cite{chen2018mobilefacenets}. All the training images are resized to 112x112, and the embedding features extracted from the face recognition module is 128-dimensional, following the original setting in MobileFaceNets. We train the face recognition model on a GTX1080 for 40 epochs, and the batchsize is 128. The recognition model achieves the accuracy of 98.3\% on LFW~\cite{LFWTech}. 

For testing, we use MeGlass~\cite{guo2018face} dataset, which is a subset of MegaFace~\cite{kemelmacher2016megaface}. In the testing set of MeGlass, there are images of 1,710 subjects. For each subject, there are 2 images with eyeglasses, one for gallery and the other for probe. There are also 2 images without glasses of each subject, still, one for gallery and the other for probe. Here we show 7 kinds of experimental protocols below. No matter which one of the protocols, all images in gallery are glasses-free faces.

\begin{itemize}
\item All images in probe are glasses-free images. 
\item All images in probe are images with glasses. 
\item All images in probe are images with glasses, but we remove the glasses with different methods including CycleGAN, SaGAN, ELEGANT, pix2pix, and our ByeGlassesGAN before face recognition.
\end{itemize}

As shown in Table~\ref{mobacc}, when images in gallery and probe are all glasses-free, the face recognition model(Experiment M) described above can achieve high accuracy on both verification and identification task. However, if we change the probe into images with glasses, the accuracy degrades a lot. 

\begin{table}[htb]
\caption{The effect of eyeglasses in face recognition: all the images in gallery are glasses-free images. The first column denotes which kind of images are there in probe. \textbf{Experiment M}: The face recognition model used is MobileFaceNet with 112x112 input image size.}
\begin{center}
\resizebox{0.8\linewidth}{!}{
\begin{tabular}{|c|c|c|c|c|}
\hline
\textbf{Experiment M} & TAR@FAR=$10^{-3}$ & TAR@FAR=$10^{-4}$ & TAR@FAR=$10^{-5}$ & Rank-1\\
\hline
no glasses & 0.9129 & 0.8567 & 0.7673 & 0.9018\\
with glasses & 0.8509 & 0.7374 & 0.5708 &0.8275\\
\hline
\end{tabular}}
\end{center}
\label{mobacc}
\end{table}

Due to the accuracy degradation for face images with eyeglasses as shown in Table~\ref{mobacc}, we then apply glasses removal methods to remove the eyeglasses in probe before face recognition. The quantitative results are shown in Table~\ref{tarfar}. As shown in row 2, row 3, row 4, and row 5 of Table~\ref{tarfar}, removing glasses with CycleGAN, SaGAN, ELEGANT, and pix2pix degrades the accuracy of face recognition. However, removing eyeglasses with our ByeGlassesGAN can improve the accuracy. Especially when FAR is small, the improvement in TAR is more evident. Comparing the unpaired training ones with our work may not be fair, but to utilize glasses removal into face recognition task, paired training is a better strategy. Besides, we also train the proposed GAN model without considering $L_{ID}$, as shown in row 7, without the Identity Extractor, the improvement of face recognition decrease since there is no mechanism to constrain the generator from producing artifacts which may be seen as noise for face recognition.\\
Here we also demonstrate face recognition experiments on our Identity Extractor used for training our ByeGlassesGAN. As shown in Table~\ref{resnet}, when FAR is $10^{-5}$, we can improve TAR even more obviously by about 6\%.

\begin{table}[htb]
\caption{Accuracy of face recognition: all the images in gallery are glasses-free images. The first column denotes the type of images with or without applying a specific glasses removal pre-processing method in the probe.}
\begin{center}
\resizebox{0.8\linewidth}{!}{
\begin{tabular}{|c|c|c|c|c|}
\hline
\textbf{Experiment M} & TAR@FAR=$10^{-3}$ & TAR@FAR=$10^{-4}$ & TAR@FAR=$10^{-5}$ & Rank-1\\
\hline
no removal  & 0.8509 & 0.7374 & 0.5708 &0.8275\\
CycleGAN & 0.8298 & 0.7205 & 0.5329 & 0.7994\\
SaGAN & 0.8386 & 0.7257 & 0.5684 & 0.8088\\
ELEGANT & 0.7497 & 0.5977 & 0.3719 & 0.6994 \\
pix2pix & 0.8444 & 0.7327 & 0.5251 & 0.8216 \\
ours without IE & 0.8573 & 0.7626 & 0.5813 & 0.8358 \\
ours & \textbf{0.8632} & \textbf{0.7719} & \textbf{0.6076} & \textbf{0.8415}\\
\hline
\end{tabular}}
\end{center}
\label{tarfar}
\end{table}

\begin{table}[htb]
\caption{Accuracy of face recognition: all the images in gallery are glasses-free images. The first column denotes the type of images with or without applying our glasses removal pre-processing method in the probe. \textbf{Experiment R}: The face recognition model used here is the Identity Extractor used for training ByeGlassesGAN with 256x256 input image size.}
\begin{center}
\resizebox{0.8\linewidth}{!}{
\begin{tabular}{|c|c|c|c|c|}
\hline
\textbf{Experiment R} & TAR@FAR=$10^{-3}$ & TAR@FAR=$10^{-4}$ & TAR@FAR=$10^{-5}$ & Rank-1\\
\hline
no removal  & 0.8801 & 0.7830 & 0.6292 & 0.8538\\
ours & \textbf{0.8836} & \textbf{0.7906} & \textbf{0.6819} & \textbf{0.8602}\\
\hline
\end{tabular}}
\end{center}
\label{resnet}
\end{table}

However, for practical applications, there might not be only glasses-portraits in probe and only glasses-free-portraits in gallery. Hence, here we do another face recognition experiment described as follows:

\begin{itemize}
    \item In gallery: 1 glasses-free image and 1 image with glasses for each person.
    \item In probe: 1 glasses-free image and 1 image with glasses for each person.
    \item For the \textit{no removal} experiment, no matter there are eyeglasses on the images or not, we use the original images.
    \item For the \textit{with removal} experiment, no matter there are eyeglasses on the images or not, we do eyeglasses removal with ByeGlassesGAN for all the images in both probe and gallery before face recognition.
\end{itemize}

As shown in Table~\ref{bothacc}, we can see applying glasses removal as a pre-processing step can still benefit face recognition even when there are glasses-free images. When FAR is $10^{-5}$, we evidently improve TAR by about 7\%. This experiment not only demonstrates the effectiveness of our glasses removal method, but also reveals that when applying our method to the glasses-free images, images remains almost the same, and the feature and identity embedding of the pre-processed face images are still well preserved.

\begin{table}[htb]
\caption{Face recognition accuracy when both gallery and probe sets contain face images with and without glasses. }
\begin{center}
\resizebox{0.8\linewidth}{!}{
\begin{tabular}{|c|c|c|c|c|}
\hline
\textbf{Experiment M} & TAR@FAR=$10^{-3}$ & TAR@FAR=$10^{-4}$ & TAR@FAR=$10^{-5}$ & Rank-1\\
\hline
no removal  & 0.8507 & 0.7516 & 0.5927 & 0.9175\\
with removal  & \textbf{0.8646} & \textbf{0.7868} & \textbf{0.6539} & \textbf{0.9289}\\
\hline
\end{tabular}}
\end{center}
\label{bothacc}
\end{table}

Besides, to make sure applying image synthesis before recognition does not harm features of faces, we have done an experiment of computing the cosine distance between features of with-glasses portrait and real glasses-free portrait of same person in the feature space of the recognition model, and the cosine distance between features of glasses-removed image and real glasses-free image. We found that our image synthesis method can effectively shorten the cosine distance for 1,335 out of 1,710 image pairs in MeGlass dataset after applying our glasses removal method. Due to the improvement of face recognition and the reduction in the cosine distance for almost 80\% image pairs, we are confident that our method cannot only manipulate visually appealing glasses-removed results, but it's also worth removing eyeglasses with our method as a preprocessing step for face recognition.

\section{Conclusions}\label{c:conclusions}

In this paper, we propose a novel multi-task framework to automatically detect the eyeglasses area and remove them from a face image. We adopt the mechanism of identity extractor to make sure the output of the proposed ByeGlassesGAN model preserves the same identity as that of the source image. As our GAN-based glasses removal framework can predict the binary mask of face shape as well, this spatial information is exploited to remove the eyeglasses from face images and achieve very realistic result.  In the face recognition experiment, we showed that our method can significantly enhance the accuracy of face recognition by about 7\%TAR@FAR=$10^{-5}$. However, there are still some limitations of our work, for example, we cannot generate convincing glasses removal results for some special glasses or when the lighting condition is very extreme. 

With the advancement of face parsing methods, we believe that combining face parsing can effectively extend this work to other attributes removal tasks, such as removing beard or hat. In the future, we will aim to make our ByeGlassesGAN more robust under special or extreme conditions, and extend the proposed framework to other face attributes removal tasks.

\clearpage
%
%
\bibliographystyle{splncs04}
\bibliography{eccv2020submissionCR}

\begin{thebibliography}{10}
\providecommand{\url}[1]{\texttt{#1}}
\providecommand{\urlprefix}{URL }
\providecommand{\doi}[1]{https://doi.org/#1}

\bibitem{bansal2016umdfaces}
Bansal, A., Nanduri, A., Castillo, C.D., Ranjan, R., Chellappa, R.: Umdfaces:
  An annotated face dataset for training deep networks. arXiv preprint
  arXiv:1611.01484v2  (2016)

\bibitem{bao2018towards}
Bao, J., Chen, D., Wen, F., Li, H., Hua, G.: Towards open-set identity
  preserving face synthesis. In: Proceedings of the IEEE Conference on Computer
  Vision and Pattern Recognition. pp. 6713--6722 (2018)

\bibitem{chen2018mobilefacenets}
Chen, S., Liu, Y., Gao, X., Han, Z.: Mobilefacenets: Efficient cnns for
  accurate real-time face verification on mobile devices. In: Chinese
  Conference on Biometric Recognition. pp. 428--438. Springer (2018)

\bibitem{choi2018stargan}
Choi, Y., Choi, M., Kim, M., Ha, J.W., Kim, S., Choo, J.: Stargan: Unified
  generative adversarial networks for multi-domain image-to-image translation.
  In: Proceedings of the IEEE Conference on Computer Vision and Pattern
  Recognition. pp. 8789--8797 (2018)

\bibitem{deng2018arcface}
Deng, J., Guo, J., Xue, N., Zafeiriou, S.: Arcface: Additive angular margin
  loss for deep face recognition. arXiv preprint arXiv:1801.07698  (2018)

\bibitem{guo2018face}
Guo, J., Zhu, X., Lei, Z., Li, S.Z.: Face synthesis for eyeglass-robust face
  recognition. In: Chinese Conference on Biometric Recognition. pp. 275--284.
  Springer (2018)

\bibitem{he2017arbitrary}
He, Z., Zuo, W., Kan, M., Shan, S., Chen, X.: Arbitrary facial attribute
  editing: Only change what you want. arXiv preprint arXiv:1711.10678
  \textbf{1}(3) (2017)

\bibitem{heusel2017gans}
Heusel, M., Ramsauer, H., Unterthiner, T., Nessler, B., Hochreiter, S.: Gans
  trained by a two time-scale update rule converge to a local nash equilibrium.
  In: Advances in Neural Information Processing Systems. pp. 6626--6637 (2017)

\bibitem{hu2019unsupervised}
Hu, B., Yang, W., Ren, M.: Unsupervised eyeglasses removal in the wild. arXiv
  preprint arXiv:1909.06989  (2019)

\bibitem{LFWTech}
Huang, G.B., Ramesh, M., Berg, T., Learned-Miller, E.: Labeled faces in the
  wild: A database for studying face recognition in unconstrained environments.
  Tech. Rep. 07-49, University of Massachusetts, Amherst (October 2007)

\bibitem{iizuka2017globally}
Iizuka, S., Simo-Serra, E., Ishikawa, H.: Globally and locally consistent image
  completion. ACM Transactions on Graphics (ToG)  \textbf{36}(4), ~107 (2017)

\bibitem{isola2017image}
Isola, P., Zhu, J.Y., Zhou, T., Efros, A.A.: Image-to-image translation with
  conditional adversarial networks. In: Proceedings of the IEEE conference on
  computer vision and pattern recognition. pp. 1125--1134 (2017)

\bibitem{jo2019sc}
Jo, Y., Park, J.: Sc-fegan: Face editing generative adversarial network with
  user's sketch and color. arXiv preprint arXiv:1902.06838  (2019)

\bibitem{kemelmacher2016megaface}
Kemelmacher-Shlizerman, I., Seitz, S.M., Miller, D., Brossard, E.: The megaface
  benchmark: 1 million faces for recognition at scale. In: Proceedings of the
  IEEE Conference on Computer Vision and Pattern Recognition. pp. 4873--4882
  (2016)

\bibitem{CelebAMask-HQ}
Lee, C.H., Liu, Z., Wu, L., Luo, P.: Maskgan: Towards diverse and interactive
  facial image manipulation. arXiv preprint arXiv:1907.11922  (2019)

\bibitem{li2016deep}
Li, M., Zuo, W., Zhang, D.: Deep identity-aware transfer of facial attributes.
  arXiv preprint arXiv:1610.05586  (2016)

\bibitem{liu2018image}
Liu, G., Reda, F.A., Shih, K.J., Wang, T.C., Tao, A., Catanzaro, B.: Image
  inpainting for irregular holes using partial convolutions. In: Proceedings of
  the European Conference on Computer Vision (ECCV). pp. 85--100 (2018)

\bibitem{liu2015deep}
Liu, Z., Luo, P., Wang, X., Tang, X.: Deep learning face attributes in the
  wild. In: Proceedings of the IEEE international conference on computer
  vision. pp. 3730--3738 (2015)

\bibitem{mao2017least}
Mao, X., Li, Q., Xie, H., Lau, R.Y., Wang, Z., Paul~Smolley, S.: Least squares
  generative adversarial networks. In: Proceedings of the IEEE International
  Conference on Computer Vision. pp. 2794--2802 (2017)

\bibitem{park2005glasses}
Park, J.S., Oh, Y.H., Ahn, S.C., Lee, S.W.: Glasses removal from facial image
  using recursive error compensation. IEEE transactions on pattern analysis and
  machine intelligence  \textbf{27}(5),  805--811 (2005)

\bibitem{pathak2016context}
Pathak, D., Krahenbuhl, P., Donahue, J., Darrell, T., Efros, A.A.: Context
  encoders: Feature learning by inpainting. In: Proceedings of the IEEE
  conference on computer vision and pattern recognition. pp. 2536--2544 (2016)

\bibitem{ronneberger2015u}
Ronneberger, O., Fischer, P., Brox, T.: U-net: Convolutional networks for
  biomedical image segmentation. In: International Conference on Medical image
  computing and computer-assisted intervention. pp. 234--241. Springer (2015)

\bibitem{upchurch2017deep}
Upchurch, P., Gardner, J., Pleiss, G., Pless, R., Snavely, N., Bala, K.,
  Weinberger, K.: Deep feature interpolation for image content changes. In:
  Proceedings of the IEEE conference on computer vision and pattern
  recognition. pp. 7064--7073 (2017)

\bibitem{wu2004automatic}
Wu, C., Liu, C., Shum, H.Y., Xy, Y.Q., Zhang, Z.: Automatic eyeglasses removal
  from face images. IEEE transactions on pattern analysis and machine
  intelligence  \textbf{26}(3),  322--336 (2004)

\bibitem{xiao2018elegant}
Xiao, T., Hong, J., Ma, J.: Elegant: Exchanging latent encodings with gan for
  transferring multiple face attributes. In: Proceedings of the European
  Conference on Computer Vision (ECCV). pp. 168--184 (2018)

\bibitem{yang2017high}
Yang, C., Lu, X., Lin, Z., Shechtman, E., Wang, O., Li, H.: High-resolution
  image inpainting using multi-scale neural patch synthesis. In: Proceedings of
  the IEEE Conference on Computer Vision and Pattern Recognition. pp.
  6721--6729 (2017)

\bibitem{yu2018bisenet}
Yu, C., Wang, J., Peng, C., Gao, C., Yu, G., Sang, N.: Bisenet: Bilateral
  segmentation network for real-time semantic segmentation. In: Proceedings of
  the European Conference on Computer Vision (ECCV). pp. 325--341 (2018)

\bibitem{yu2018free}
Yu, J., Lin, Z., Yang, J., Shen, X., Lu, X., Huang, T.S.: Free-form image
  inpainting with gated convolution. arXiv preprint arXiv:1806.03589  (2018)

\bibitem{yu2018generative}
Yu, J., Lin, Z., Yang, J., Shen, X., Lu, X., Huang, T.S.: Generative image
  inpainting with contextual attention. arXiv preprint arXiv:1801.07892  (2018)

\bibitem{zhang2018generative}
Zhang, G., Kan, M., Shan, S., Chen, X.: Generative adversarial network with
  spatial attention for face attribute editing. In: Proceedings of the European
  Conference on Computer Vision (ECCV). pp. 417--432 (2018)

\bibitem{zhu2017unpaired}
Zhu, J.Y., Park, T., Isola, P., Efros, A.A.: Unpaired image-to-image
  translation using cycle-consistent adversarial networks. In: Proceedings of
  the IEEE International Conference on Computer Vision. pp. 2223--2232 (2017)

\end{thebibliography}
\end{document}